\documentclass[12pt]{article}
\usepackage{epsfig}
\def\be{\begin{equation}}
\def\ee{\end{equation}}
\def\bea{\begin{eqnarray}}
\def\eea{\end{eqnarray}}
\usepackage{graphicx}

\catcode`\@=11
\def\lsim{\mathrel{\mathpalette\@versim<}}
\def\gsim{\mathrel{\mathpalette\@versim>}}
\def\@versim#1#2{\vcenter{\offinterlineskip
\ialign{$\m@th#1\hfil##\hfil$\crcr#2\crcr\sim\crcr } }}
\catcode`\@=12

\catcode`@=12
\begin{document}
\thispagestyle{empty}
\begin{flushright}
UCRHEP-T542\\
March 2014\
\end{flushright}
\vspace{0.6in}
\begin{center}
{\LARGE \bf Scotogenic $Z_2$ or $U(1)_D$ Model of\\ Neutrino Mass  
with $\Delta(27)$ Symmetry\\}
\vspace{1.2in}
{\bf Ernest Ma and Alexander Natale\\}
\vspace{0.2in}
{\sl Department of Physics and Astronomy, University of California,\\
Riverside, California 92521, USA\\}
\end{center}
\vspace{1.2in}
\begin{abstract}\
The scotogenic model of radiative neutrino mass with $Z_2$ or $U(1)_D$ dark 
matter is shown to accommodate $\Delta(27)$ symmetry naturally.  The 
resulting neutrino mass matrix is identical to either of two forms, 
one proposed in 2006, the other in 2008.  These two structures are studied 
in the context of present neutrino data, with predictions of $CP$ violation 
and neutrinoless double beta decay.
\end{abstract}

\newpage
\baselineskip 24pt

To understand the pattern of neutrino mixing, non-Abelian discrete symmetries 
have been used frequently in the past several years, starting with 
$A_4$~\cite{mr01,m02,bmv03,m04,af05,bh05}.  Another symmetry $\Delta(27)$ 
was also studied~\cite{m06,dkr07,m08} some years ago.  Using the fact 
that it admits geometric $CP$ violation~\cite{bgg84}, it has been 
proposed recently for understanding the $CP$ phases in the mixing of 
quarks~\cite{bdl12,dp13} and of leptons~\cite{m13,dp13x}.

In a parallel development, there is a large body of literature on the 
radiative generation of neutrino mass through dark matter.  The simplest 
original (scotogenic) one-loop model~\cite{m06x} adds one extra scalar 
doublet $(\eta^+,\eta^0)$ and three neutral fermion singlets $N_{1,2,3}$ 
together with an exactly conserved $Z_2$ symmetry under which the new 
particles are odd and the standard-model (SM) particles are even.  
The resulting one-loop diagram for Majorana neutrino mass is shown 
in Fig.~1.
\begin{figure}[htb]
\vspace*{-3cm}
\hspace*{-3cm}
\includegraphics[scale=1.0]{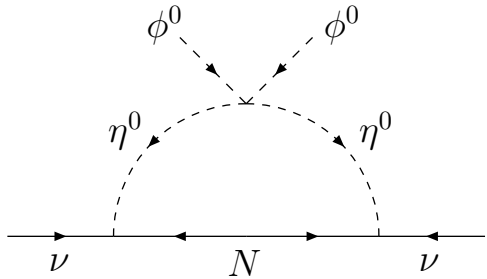}
\vspace*{-21.5cm}
\caption{One-loop generation of neutrino mass with $Z_2$ symmetry.}
\end{figure}

A variation of this mechanism was recently proposed~\cite{mpr13}, using two 
extra scalar doublets $(\eta^+_{1,2},\eta^0_{1,2})$ transforming as $\pm 1$ 
under an $U(1)_D$ gauge symmetry together with three Dirac fermion singlets 
$N_{1,2,3}$ transforming as $+1$.  The resulting one-loop diagram for 
Majorana neutrino mass is shown in Fig.~2.
\begin{figure}[htb]
\vspace*{-3cm}
\hspace*{-3cm}
\includegraphics[scale=1.0]{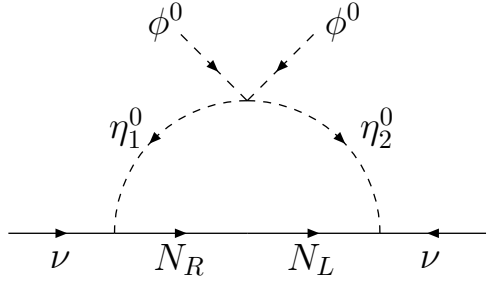}
\vspace*{-21.5cm}
\caption{One-loop generation of neutrino mass with $U(1)_D$ symmetry.}
\end{figure}

Combining these two ideas, it is shown in this paper that $\Delta(27)$ is 
naturally adapted to realize the two neutrino mass matrices proposed 
earlier~\cite{m06,m08} without additional particle content in the loop, 
in either the $Z_2$ or $U(1)_D$ case.  We then study their implications 
in the context of present neutrino data.

The group $\Delta(27)$ has nine one-dimensional representations 
$\underline{1}_i (i = 1,...,9)$ and two three-dimensional ones 
$\underline{3}, \underline{3}^*$.  Their multiplication rules are
\begin{equation}
\underline{3} \times \underline{3}^* = \sum_{i=1}^9 \underline{1}_i, ~~~~~ 
\underline{3} \times \underline{3} = \underline{3}^* + \underline{3}^* + 
\underline{3}^*.
\end{equation}
In the decomposition of $\underline{3} \times \underline{3} \times 
\underline{3}$, there are three invariants: $111+222+333$ and 
$123+231+312 \pm (132+213+321)$.  In Fig.~1, let $\Phi, \eta \sim 
\underline{1}_1$, $\nu \sim \underline{3}$, $N \sim \underline{3}^*$, 
then the $3 \times 3$ Majorana neutrino mass matrix is proportional 
to the $3 \times 3$ Majorana $N$ mass matrix, which is nonzero from the 
vacuum expectation values of a neutral scalar $\zeta \sim \underline{3}$ 
under $\Delta(27)$ with the Yukawa couplings $f_{ijk} N_i N_j \zeta_k^*$. 
In Fig.~2, let $\Phi, \eta_{1,2} \sim \underline{1}_1$, $\nu \sim 
\underline{3}$, $N_R \sim \underline{3}$, $N_L \sim \underline{3}^*$, 
then the same result is obtained with $f_{ijk} \bar{N}_{Li} N_{Rj} \zeta_k$. 
In both cases, the neutrino mass matrix is of the form
\begin{equation}
{\cal M}_\nu = \pmatrix{fa & c & b \cr c & fb & a \cr b & a & fc},
\end{equation}
where $a,b,c$ are proportional to the three arbitrary vacuum expectation 
values of $\zeta$.

In Ref.~\cite{m06}, with $l^c \sim \underline{3}^*$ and $\phi_{1,2,3} \sim 
\underline{1}_{1,2,3}$, the charged-lepton mass matrix is diagonal, whereas in 
Ref.~\cite{m08}, with $l^c \sim \underline{3}$ and $\phi \sim \underline{3}$, 
it is given by
\begin{equation}
{\cal M}_l = U_\omega \pmatrix{m_e & 0 & 0 \cr 0 & m_\mu & 0 \cr 0 & 0 
& m_\tau} U_\omega^\dagger,
\end{equation}
where $U_\omega$ is the familiar
\begin{equation}
U_\omega = {1 \over \sqrt{3}} \pmatrix{1 & 1 & 1 \cr 1 & \omega & \omega^2 
\cr 1 & \omega^2 & \omega},
\end{equation}
with $\omega = \exp(2\pi i/3) = -1/2 + i \sqrt{3}/2$.  Both models are 
consistent with $\theta_{13} \neq 0$, but they were proposed before 
its determination in 2012.

Consider first the case where the charged-lepton mass matrix is 
diagonal.  In Ref.~\cite{m06}, two solutions were found with $\theta_{13}=0$; 
one with $f \simeq 1$, the other with $f \simeq -0.5$.  The former turns 
out to be unacceptable because $\theta_{13}$ is always very small.  The 
latter has a solution as shown below.  Let $f = -0.5 + \epsilon$, 
$a = b (1 + \eta)$ and $c = b (1 - \delta)$, then in the tribimaximal 
basis, the neutrino mass matrix becomes
\begin{equation}
{\cal M}_\nu^{TB} = \pmatrix{-{3 \over 2} + \epsilon + {3 \over 4} \delta 
& -{(2 \eta + \delta) \over 2 \sqrt{2}} & {\sqrt{3} \over 4} \delta \cr 
-{(2 \eta + \delta) \over 2 \sqrt{2}} & {3 \over 2} + \epsilon + {1 \over 2} 
\eta - {1 \over 2} \delta & {\sqrt{3} \over 2 \sqrt{2}} \delta \cr 
{\sqrt{3} \over 4} \delta & {\sqrt{3} \over 2 \sqrt{2}} \delta & 
-{3 \over 2} + \epsilon - \eta + {1 \over 4} \delta} b,
\end{equation}
where $\epsilon, \eta, \delta$ are all assumed to be small compared to one.
We define $\delta + 2\eta = \zeta$ and assume all parameters tio be real, 
then
\begin{eqnarray}
\Delta m^2_{21} \simeq {3 \over 4} (8 \epsilon + \zeta) b^2, ~~~ 
\Delta m^2_{31} \simeq {3 \over 2} \zeta b^2, ~~~ \sin \theta_{13} \simeq 
\pm {\delta \over \sqrt{2} \zeta}, ~~~ \tan \theta_{12} \simeq 
{1 \over \sqrt{2}} 
\left[ {1 - \zeta/6 \over 1 + \zeta/12} \right].
\end{eqnarray}
Using $\tan^2 \theta_{12} = 0.45$, we find $\zeta = 0.209$.  Hence 
$\Delta m^2_{31} > 0$, i.e. normal ordering of neutrino masses. 
Using $\Delta m^2_{31} = 2.32 \times 10^{-3}$ eV$^2$, we find $b = 0.086$ eV. 
Using $\sin \theta_{13} = \pm 0.16$, we find $\delta = \pm 0.047$. 
Using $\Delta m^2_{21} = 7.50 \times 10^{-5}$ eV$^2$, we find 
$8 \epsilon + \zeta = 0.0135$.  This predicts $\sin^2 2 \theta_{23} = 
0.966$ and $m_{ee} = |f a| = 0.05$ eV for the effective Majorana neutrino 
mass in neutrinoless double beta decay..

We consider also the case with $\delta$ purely imaginary, in which case 
$\sin^2 2 \theta_{23} = 1$ is guaranteed in the limit of a symmetry based 
on a generalized $CP$ transformation~\cite{gl04}.  Using a 
complete numerical analysis, we plot in Fig.~3 the predictions of this 
model for $m_{ee}$ as a function of $\sin^2 2 \theta_{12}$ for 
$\sin^2 2 \theta_{13} = 0.095 \pm 0.010$.  The higher (lower) band corresponds 
to $\delta$ real (purely imaginary), with the allowed region in between for 
any arbitrary phase.
\begin{figure}[htb]
\hspace*{1cm}
\includegraphics[scale=1.3]{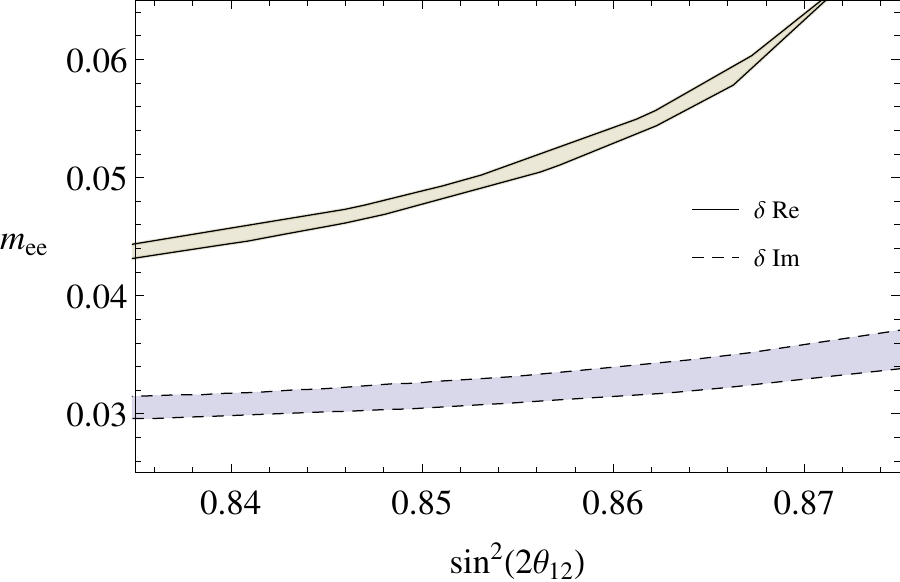}
\caption{Predictions of $m_{ee}$ versus $\sin^2 2 \theta_{12}$ for 
$\sin^2 2 \theta_{13} = 0.095 \pm 0.010$.}
\end{figure}

We plot in Fig.~4 the invariant $J_{CP}$ as a function of $\sin^2 2 \theta_{13}$ 
for $\delta$ purely imaginary and $\sin^2 \theta_{12} = 0.857 \pm 0.024$.
\begin{figure}[htb]
\hspace*{1cm}
\includegraphics[scale=1.0]{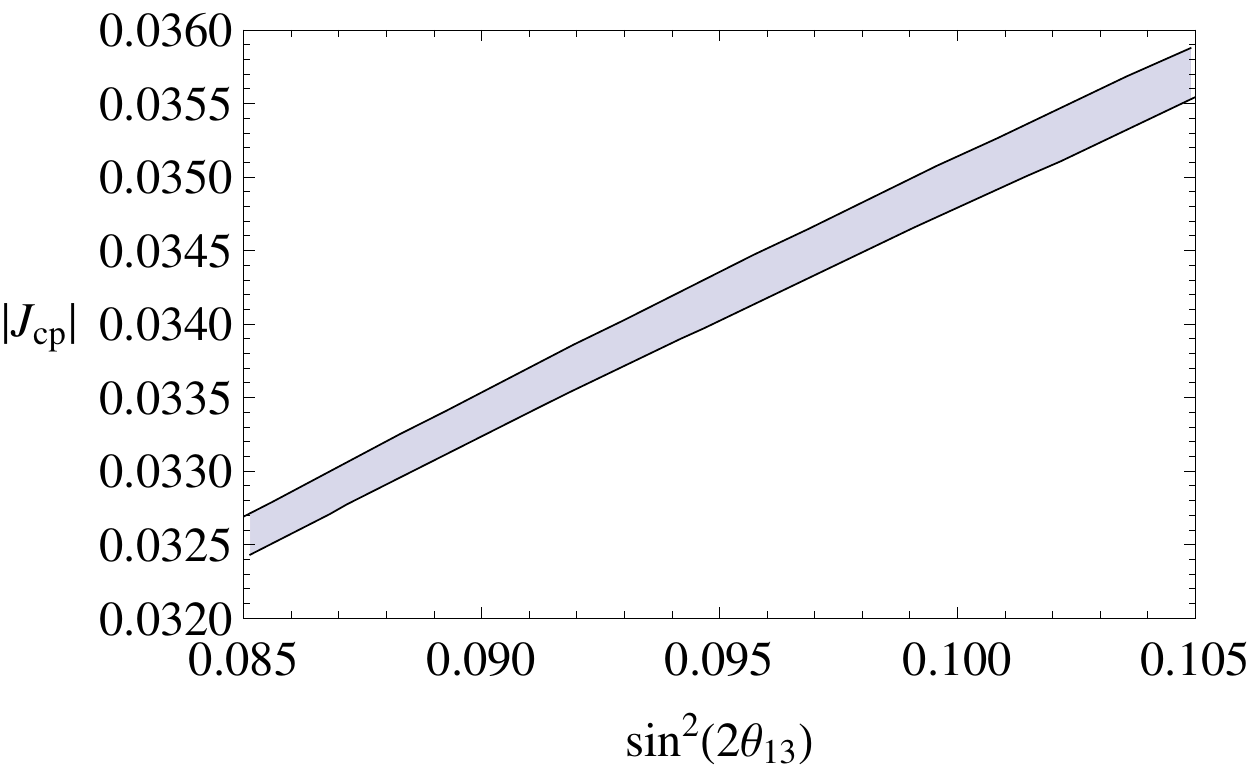}
\caption{Predictions of $|J_{CP}|$ versus $\sin^2 2 \theta_{13}$ for 
$\delta$ purely imaginary and $\sin^2 2 \theta_{12} = 0.857 \pm 0.024$.}
\end{figure}

Consider next the case where ${\cal M}_l$ is given by Eq.~(3).  In the 
tribimaximal basis,
\begin{equation}
{\cal M}_\nu^{TB} = \pmatrix{a + f(b+c)/2 & (b+c)/\sqrt{2} & f(-b+c)/2 \cr 
(b+c)/\sqrt{2} & fa & (b-c)/\sqrt{2} \cr f(-b+c)/2 & (b-c)/\sqrt{2} & 
a - f(b+c)/2}.
\end{equation}
Let $f = -1 + \epsilon'$, $\eta' = (b+c)/2a$, $\delta' = (b-c)/2a$, then
\begin{equation}
{\cal M}_\nu^{TB} \simeq \pmatrix{1-\eta' & \sqrt{2} \eta' & \sqrt{2} \delta' 
\cr \sqrt{2} \eta' & -1+\epsilon' & \delta' \cr \sqrt{2} \delta' & \delta' 
& 1+\eta'} a,
\end{equation}
where $\epsilon', \eta', \delta'$ are all assumed to be small 
compared with one.
This turns out to have the same approximate solution as Eq.~(5) with the 
following substitutions:
\begin{equation}
a = {-3b \over 2}, ~~ \eta' = {\zeta \over 6}, ~~ \delta' = {\delta \over 
2\sqrt{6}}, ~~ \epsilon' = {-4\epsilon \over 3}.
\end{equation}
The predicted $m_{ee}$ is also approximately the same.  Thus the physical 
manifestations of this second model are indistinguiable from those of 
the first to a good approximation.

\noindent \underline{Acknowledgment}~:~This work is supported in part 
by the U.~S.~Department of Energy under Grant No.~DE-SC0008541.

\bibliographystyle{unsrt}

\end{document}